\documentclass[traditabstract]{aa} 
\usepackage{graphicx}
\usepackage{txfonts}
\begin{document}
   \title{The nature of the late achromatic bump in GRB\,120326A}
   
   \author{A. Melandri\inst{1}, F. J. Virgili\inst{2}, C. Guidorzi\inst{3}, M. G. Bernardini\inst{1}, S. Kobayashi\inst{2}, C. G. Mundell\inst{2}, A. Gomboc\inst{4}, B. Dintinjana\inst{5,4}, V.-P. Hentunen\inst{6}, J. Japelj\inst{4}, D. Kopa\v{c}\inst{4,2}, D. Kuroda\inst{7}, A. N. Morgan\inst{8}, I. A. Steele\inst{2}, U. Quadri\inst{9}, G. Arici\inst{10}, D. Arnold\inst{2}, R. Girelli\inst{9}, H. Hanayama\inst{11}, N. Kawai\inst{12}, H. Miku\v{z}\inst{5,4}, M. Nissinen\inst{6}, T. Salmi\inst{6}, R. J. Smith\inst{2}, L. Strabla\inst{9}, M. Tonincelli\inst{10}, A. Quadri\inst{9}}

   \institute{
   $^{1}$ INAF - Osservatorio Astronomico Brera, Via E. Bianchi 46, I-23807, Merate (LC), Italy\\
              \email{andrea.melandri@brera.inaf.it}\\
   $^{2}$ ARI - Liverpool John Moores University, IC2 Liverpool Science Park, 146 Brownlow Hill, Liverpool, L3 5RF, UK\\
   $^{3}$ Dipartimento di Fisica e Scienza della Terra, Via Saragat 1, I-44122, Ferrara, Italy\\
   $^{4}$ Faculty of Mathematics and Physics, University of Ljubljana, Jadranska 19, SI-1000 Ljubljana, Slovenia\\
   $^{5}$ \v{C}rni Vrh Observatory, Predgri\v{z}e 29A, SI-5274, \v{C}rni Vrh nad Idrijo, Slovenia\\
   $^{6}$ Taurus Hill Observatory, H\"ark\"am\" aentie 88, 79480 Kangaslampi, Finland\\
   $^{7}$ Okayama Astrophysical Observatory - NAOJ, 3037-5 Honjo, Kamogata, Asakuchi, Okayama, 719-0232, Japan\\
   $^{8}$ Department of Astronomy, University of California, Berkeley, CA 94720-3411, USA\\
   $^{9}$ Osservatorio Astronomico Bassano Bresciano, Via San Michele 4, I-25020, Bassano Bresciano (BS), Italy\\
   $^{10}$ Osservatorio Astronomico di Cima Rest, Via Rest, I-25080, Magasa (BS), Italy\\
   $^{11}$ Ishigakijima Astronomical Observatory - NAOJ, 1024-1, Arakawa, Ishigaki, Okinawa 907-0024, Japan\\
   $^{12}$ Department of Physics, Tokyo Institute of Technology, 2-12-1 Ookayama, Meguro-ku, Tokyo 152-8551, Japan\\
             }



 \abstract{The long {\it Swift} gamma-ray burst GRB\,120326A at redshift $z=1.798$ exhibited a multi-band light curve with a striking feature: a late-time, long-lasting achromatic rebrightening, rarely seen in such events. Peaking in optical and X-ray bands $\sim 35$~ks ($\sim 12.5$~ks in the GRB rest frame) after the 70-s GRB prompt burst, the feature brightens nearly two orders of magnitude above the underlying optical power-law decay.  Modelling the multiwavelength light curves, we investigate possible causes of the rebrightening in the context of the standard fireball model. We exclude a range of scenarios for the origin of this feature: reverse-shock flash, late-time forward shock peak due to the passage of the maximal synchrotron frequency through the optical band, late central engine optical/X-ray flares, interaction between the expanding blast wave and a density enhancement in the circumburst medium and gravitational microlensing. Instead we conclude that the achromatic rebrightening may be caused by a refreshed forward shock or a geometrical effect.  In addition, we identify an additional component after the end of the prompt emission, that shapes the observed X-ray and optical light curves differently, ruling out a single overall emission component to explain the observed early time emission.}


\authorrunning{A. Melandri et al. 2014}
\titlerunning{The nature of GRB\,120326A late time re-brightening}

\maketitle

%

\section{Introduction}

Gamma-Ray Bursts (GRBs) are brief and intense pulses of $\gamma$-rays (prompt emission) followed by long-lasting afterglow emission that can span the entire electromagnetic spectrum from X-rays to radio bands. Since the advent of the {\it Swift} satellite \cite{gehrels} X-ray afterglows have been unarguably the most densely sampled for the majority of GRBs, from very early times until days/weeks after the burst event. This led, in the context of the standard fireball model, to the definition of a canonical light curve in the X-ray band \cite{nousek} comprising: 1) an initial steep decay (possibly reminiscent of the high-latitude prompt emission) lasting until $\sim 10^{2}$~s; 2) a possible shallow (or rising) phase (defined "plateau" at large) between $\sim 10^{2} - 10^{4}$~s that might be due to prolonged central engine activity, energy injection into the forward shock or variation of microphysical parameters; 3) a normal phase up to $\sim 10^{5}$~s showing the decaying afterglow emission of the forward shock interacting with the external medium; 4) a late phase with a steeper decay, not always seen, consistent with a jet break. In up to 50\% of GRBs, flare activity due to internal shocks is seen superimposed on the first two phases \cite{zhang2, burrows}.

At longer wavelengths, the behaviour of the light curves may not always follow that observed in the X-ray. If the same light curve features are present contemporaneously in different bands, the behaviour is described as achromatic and the radiation is interpreted as having been produced by a single emission mechanism or the achromatic behaviour is due to geometric effect, e.g. late-time steepening of the light curves hours to days after the GRB, due to a jet break.

In contrast, early time emission is often chromatic as the typical synchrotron frequencies pass through the observing bands. In addition, multiple emission components from different locations in the relativistic outflow may be temporally superimposed in the observed light curve. The key components of the early afterglow are expected to be emission from the external reverse and forward shocks \cite{jure}, with the possibility of additional flares or re-brightenings due to energy injection from long-lived central engine activity \cite{mela3, virgili}, or interaction between the advancing shock and inhomogenities in the circumburst medium \cite{mundell}.  In a small number of GRBs with optical emission observed contemporaneously with the prompt $\gamma$-ray emission, rapid variability ($\Delta$t/t $\lesssim 1$) and steep rise/decay indices suggest an internal dissipation for the origin of the optical emission \cite{monfardini, kopac}. 

Overall, chromaticity is more often seen in gamma-ray burst afterglow light curves, with only a minority of bursts clearly showing similar behaviour in the X-ray and optical bands \cite{pana1, ghise, mela1, nardini2, guidorzi2, mela2}. In those bursts, with high quality data, the similarity between the X-ray and optical bands is striking (GRB100901A, \cite{gomboc}; GRB 081028,  \cite{raffa}), while for others the evidence for achromaticity is only marginal (GRB 071010A, \cite{covino1}; GRB 091029, \cite{filgas}). 

Here we present high quality, panchromatic observations from X-ray to radio bands of GRB\,120326A, which exhibits an unusual and pronounced late-time achromatic bump occurring simultaneously in X-ray and optical bands between $10^{3} - 10^{5}$~s in the rest-frame of the burst. We test a wide range of different scenarios to explain the re-brightening: reverse-shock emission, the passage of the typical frequency, the onset of the afterglow, a refreshed shock, a late-time flare, a density enhancement of the ambient medium, a geometrical effect and gravitational microlensing. Throughout the paper we assume a standard cosmology with $H_{\rm 0}$ = 72 km s$^{-1}$ Mpc$^{-1}$, $\Omega_{\rm m}$ = 0.27, and $\Omega_{\rm \Lambda}$ = 0.73. The respective temporal and spectral decay indices $\alpha$ and $\beta$ are defined by f$_{\nu}$(t) $\propto$ t$^{-\alpha} \nu^{-\beta}$ and unless stated otherwise, errors are statistical only.



   \begin{figure*}
   \centering
   \includegraphics[width=6.0cm,height=8.5cm,angle=270]{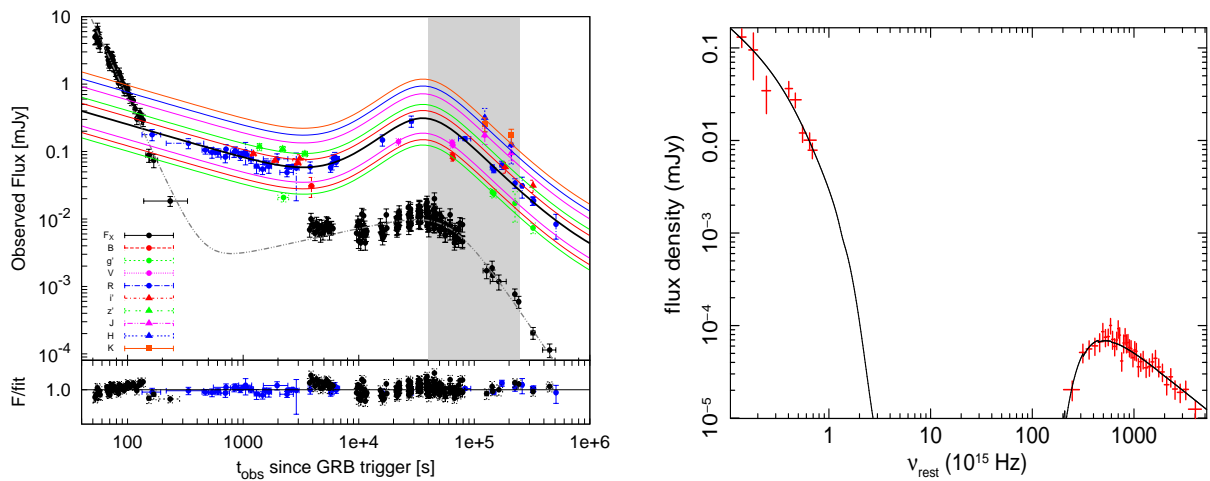}
  \includegraphics[width=6.0cm,height=7.5cm,angle=270]{restfr_sed_mJy_1e15Hz_1dayrf.ps}   
   \caption{{\it Left panel}: Panchromatic light curve of GRB\,120326A in the observer frame. In the X-ray band we draw only the function of the final fit (dot-dashed grey line), while for the optical bands we show the final fit for the $R$ filter (black solid line) and also the re-scaled fit for each wavelength (coloured lines). The scaling factors for different filters are 0.48, 0.4, 0.6, 1.3, 1.6, 2.3, 3.0 and 3.8 for the $Bg'Vi'z'JHK$ bands, respectively. The shaded grey region refers to the time interval over which the XRT spectrum has been extracted. The lower panels show the residuals of the optical and X-ray fits. {\it Right panel}: Spectral energy distribution at the reference time t$^{\rm SED}_{\rm rf}$ = 1 d (see main text for details).}
              \label{FigLC}%
    \end{figure*}

\section{Observations}

On 2012 March 26 at 01:20:29 UT (= T$_{0}$), the {\it Swift}/BAT triggered on the long GRB\,120326A \cite{siegel}. The BAT light curve showed two well defined precursor peaks (each $\sim 30$~s wide) followed by a main FRED (Fast Rise Exponential Decay) peak that returned to the background level after $\sim 20$~s. The total duration of the burst was $T_{90} = 69 \pm 8$~s \cite{barth1}. 

{\it Swift}/XRT promptly detected the afterglow emission in the X-ray with good precision within $\sim 1$~min after the event, whereas {\it Swift}/UVOT was not able to detect any credible candidate in the optical bands during its first observations. The optical afterglow was detected few minutes later from the ground by small (TAROT; \cite{klotz1}) and large (Liverpool Telescope; \cite{guidorzi1}) robotic telescopes, identifying a slowly decaying counterpart.

Spectroscopic observations performed with the 10.4m GTC telescope $\sim 2$ hr after the event, showed several absorption features at a common redshift of $z = 1.798$ \cite{tello}. This event was also detected by Fermi-GBM and displayed an average $\gamma$-ray fluence of $\sim 3.5 \times 10^{-6}$ erg~cm$^{-2}$ in the 10-1000 keV band with a peak energy E$_{\rm peak}$ = 46 $\pm$ 4 keV \cite{collazzi}. The redshift of the burst (corresponding to a luminosity distance of $\sim 1.37 \times 10^{4}$~Mpc) resulted in an isotropic energy estimate of E$_{\rm \gamma,iso} = (3.45 \pm 0.14) \times 10^{52}$~erg in the rest-frame [1-10000] keV bandpass.

\subsection{Optical and Near-infrared data}

The optical afterglow reported by Klotz et al. (2012a) was observed by many telescopes in the subsequent couple of days. We acquired images in the optical bands ($g'Ri'z'$), starting from $\sim 3.5$ min after the burst, with the 2-m Liverpool robotic telescope \cite{guidorzi1}, the 0.6-m Cichocki robotic telescope at the \v{C}rni Vrh Observatory \cite{bojan}, the 0.43-m  T17 telescope \cite{hent}, the 0.32-m robotic telescope at the Bassano-Bresciano Observatory \cite{quadri1, quadri2, quadri3, quadri4, quadri5}, the MITSuME 1.05-m telescope at the Ishigakijima Observatory \cite{kuroda2, kuroda3} and the 0.51-m telescope at the Cima Rest Observatory \cite{toni}. Near-infrared data were acquired with the 1.3-m Pairitel telescope in the $JHKs$-bands \cite{morgan}. A summary of our observations is given in Table \ref{logmag}.

\subsection{Radio, mm and sub-mm data}

The afterglow of GRB~120326A was detected by the Sub-Millimeter Array (SMA; \cite{urata, urata2}) at the typical frequency of $\nu_{\rm SMA} = 219$ GHz at a flux density of $f_{\rm SMA} = 3.1 \pm 0.5$ mJy, by the Combined Array for Research in Millimeter-Wave Astronomy (CARMA; \cite{perley}) at $\nu_{\rm CARMA} = 92.5$ GHz at a flux density of $f_{\rm CARMA} = 3.2 \pm 0.4$ mJy and by the Expanded Very Large Array (EVLA; \cite{laskar}) at $\nu_{\rm EVLA} = 21.9$ GHz at a flux density of $f_{\rm EVLA} \sim 1.36$ mJy.

 \begin{table*}
 \centering
 \caption[]{Optical observations of GRB\,120326A. Columns are: beginning time ($\Delta$t), length (t$_{\rm exp}$) of the exposure, optical filter used for the observations, magnitudes (with errors) and reference to the telescope used for each observation. Magnitudes have not been corrected for Galactic absorption along the line of sight (E$_{\rm (B-V)} = 0.05$; \cite{sf}). References for data taken from GCNs are: (1) \cite{klotz2}; (2) \cite{skynet}; (3) \cite{walker}; (4) \cite{kuin}; (5) \cite{gmg}; (6) \cite{jang}.}
 \label{logmag}
 \begin{tabular}{ccccc|ccccc}
 \hline
$\Delta$t & t$_{\rm exp}$ & Filter & Magnitude & Ref. & $\Delta$t & t$_{\rm exp}$ & Filter & Magnitude & Ref. \\
\hline
[s] & [s] & & & & [s] & [s] & & & \\
\hline
3697 & 393 & $B$ & 20.23 $\pm$ 0.16 & (4)               & 2715 & 330 & $R$ & 19.44 $\pm$ 0.67 & LT-SkyCam \\                 
64440 & 300 & $B$ & 19.40 $\pm$ 0.10 & (5) 		& 3771 & 60 & $R$ & 19.40 $\pm$ 0.20 & (3) \\                        
65520 & 300 & $B$ & 19.40 $\pm$ 0.10 & (5)		& 5659 & 120 & $R$ & 19.42 $\pm$ 0.16 & T17 \\                       
21959 & 295 & $V$ & 18.67 $\pm$ 0.12 & (4)		& 5822 & 120 & $R$ & 19.27 $\pm$ 0.21 & T17 \\                       
64080 & 300 & $V$ & 18.7 $\pm$ 0.10 & (5)		& 5985 & 120 & $R$ & 19.08 $\pm$ 0.14 & T17 \\                       
65160 & 300 & $V$ & 18.8 $\pm$ 0.10 & (5)		& 6149 & 120 & $R$ & 19.02 $\pm$ 0.17 & T17\\                        
1994 & 480 & SDSS-$g'$ & 20.77 $\pm$ 0.15 & (2)		& 6311 & 120 & $R$ & 19.22 $\pm$ 0.15 & T17\\                        
142799 & 1200 & SDSS-$g'$ & 20.58 $\pm$ 0.14 & MITSuME 	& 6475 & 120 & $R$ & 19.10 $\pm$ 0.09 & T17  \\                      
146890 & 1140 & SDSS-$g'$ & 20.62 $\pm$ 0.15 & MITSuME 	& 15996 & 60 & $R$ & 18.40 $\pm$ 0.20 & (3) \\                       
223766 & 2280 & SDSS-$g'$ & 21.01 $\pm$ 0.32 & MITSuME 	& 27640 & 300 & $R$ & 17.63 $\pm$ 0.20 & (6) \\                      
317981 & 8700 & SDSS-$g'$ & 21.90 $\pm$ 0.19 & MITSuME	& 72987 & 13080 & $R$ & 18.35 $\pm$ 0.07 & Bassano Obs.\\            
133 & 60 & $R$ & 18.20 $\pm$ 0.20 & (1)			& 142799 & 1200 & $R$ & 19.41 $\pm$ 0.09 & MITSuME  \\               
216 & 240 & $R$ & 18.52 $\pm$ 0.18 & LT-RINGO2         	& 146890 & 1140 & $R$ & 19.52 $\pm$ 0.09 & MITSuME \\                
441 & 60 & $R$ & 18.77 $\pm$ 0.14 & \v{C}rni Vrh Obs.      	& 166916 & 13080 & $R$ & 19.33 $\pm$ 0.15 & Bassano Obs. \\          
508 & 60 & $R$ & 18.86 $\pm$ 0.19 & \v{C}rni Vrh Obs.      	& 223766 & 2280 & $R$ & 20.00 $\pm$ 0.17 & MITSuME\\                 
552 & 320 & $R$ & 18.74 $\pm$ 0.18 & LT-RINGO2         	& 253037 & 13080 & $R$ & 20.10 $\pm$ 0.31 & Bassano Obs. \\          
575 & 60 & $R$ & 18.87 $\pm$ 0.15 & \v{C}rni Vrh Obs.      	& 317981& 4320 & $R$ & 20.66 $\pm$ 0.15 & MITSuME \\                 
642 & 120 & $R$ & 19.03 $\pm$ 0.19 & \v{C}rni Vrh Obs.     	& 323443 & 4380 & $R$ & 20.65 $\pm$ 0.14 & MITSuME \\                
776 & 120 & $R$ & 18.86 $\pm$ 0.17 & \v{C}rni Vrh Obs.     	& 505266 & 12000 & $R$ & 21.52 $\pm$ 0.38 & Cima Rest Obs.\\         
841 & 30 & SDSS-$r'$ & 18.81 $\pm$ 0.16 & LT-RATCam    	& 1165 & 120 & SDSS-$i'$ & 19.06 $\pm$ 0.07 & LT-RATCam\\            
910 & 180 & $R$ & 18.92 $\pm$ 0.11 & \v{C}rni Vrh Obs.	& 1775 & 120 & SDSS-$i'$ & 19.31 $\pm$ 0.06 & LT-RATCam  \\          
915 & 270 & $R$ & 18.83 $\pm$ 0.38 & LT-SkyCam   	& 2793 & 360 & SDSS-$i'$ & 19.40 $\pm$ 0.07 & LT-RATCam \\           
931 & 30 & SDSS-$r'$ & 19.00 $\pm$ 0.14 & LT-RATCam    	& 3089 & 800 & SDSS-$i'$ & 19.24 $\pm$ 0.15 & (2) \\                 
1005 & 120 & SDSS-$r'$ & 18.95 $\pm$ 0.04 & LT-RATCam  	& 183282 & 4620 & I & 19.55 $\pm$ 0.20 & MITSuME\\                   
1111 & 120 & $R$ & 19.10 $\pm$ 0.15 & \v{C}rni Vrh Obs.    	& 317982 & 8700 & I & 20.24 $\pm$ 0.22 & MITSuME\\                   
1245 & 120 & $R$ & 19.37 $\pm$ 0.24 & \v{C}rni Vrh Obs.    	& 1326 & 120 & SDSS-$z'$ & 18.79 $\pm$ 0.13 & LT-RATCam  \\          
1380 & 180 & $R$ & 19.48 $\pm$ 0.18 & \v{C}rni Vrh Obs.    	& 2075 & 240 & SDSS-$z'$ & 18.87 $\pm$ 0.10 & LT-RATCam\\            
1478 & 120 & SDSS-$r'$ & 19.25 $\pm$ 0.05 & LT-RATCam 	& 3241 & 360 & SDSS-$z'$ & 19.05 $\pm$ 0.10 & LT-RATCam\\            
1495 & 945 & $R$ & 19.10 $\pm$ 0.20 & (1)		& 120852 & 4727 & J & 17.8 $\pm$ 0.4 & PAIRITEL\\                    
1581 & 180 & $R$ & 19.38 $\pm$ 0.22 & \v{C}rni Vrh Obs.    	& 206964 & 4048 & J & 18.9 $\pm$ 0.5 & PAIRITEL\\                    
1610 & 120 & SDSS-$r'$ & 19.33 $\pm$ 0.06 & LT-RATCam	& 120852 & 4727 & H & 16.3 $\pm$ 0.4 & PAIRITEL \\                   
2076 & 640 & SDSS-$r'$ & 19.87 $\pm$ 0.15 & (2)		&  206964 & 4048 & H & 17.3 $\pm$ 0.5 & PAIRITEL\\                   
2363 & 120 & SDSS-$r'$ & 19.40 $\pm$ 0.06 & LT-RATCam  	& 120852 & 4727 & K & 16.03 $\pm$ 0.25 & PAIRITEL \\                 
2495 & 120 & SDSS-$r'$ & 19.40 $\pm$ 0.05 & LT-RATCam  	& 206964 & 4048 & K & 16.45 $\pm$ 0.25 & PAIRITEL \\                 
2627 & 120 & SDSS-$r'$ & 19.38 $\pm$ 0.06 & LT-RATCam	& & & & & \\                                                         
\hline
\end{tabular}
\end{table*}

\section{Results}

We calibrated our optical images with respect to several field stars. In particular, we calibrated SDSS-$r'$ images acquired with LT with respect to the $R$ band in order to have a well sampled $R$-band light curve from early to late times. Then we converted the observed optical-NIR magnitudes (Table \ref{logmag}) into flux densities \cite{fuku} after taking into account the Galactic extinction (E$_{\rm (B-V)} = 0.05$, $A_{B} = 0.182$~mag, $A_{g'} = 0.169$~mag, $A_{V} = 0.134$~mag, $A_{R} = 0.106$~mag, $A_{i'} = 0.074$~mag, $A_{z'} = 0.063$~mag, $A_{J} = 0.035$~mag, $A_{H} = 0.022$~mag, $A_{K} = 0.015$~mag; \cite{sf}).

\subsection{Light curve}

In Fig.\ref{FigLC} (left panel) we report the optical and X-ray light curves of GRB\,120326A. We fit the X-ray and best sampled optical band ($R$ filter) independently, in the time interval [10-10$^6$]~s, with the same number of components: a single power-law decay (PL) plus smoothly broken-power law (BPL, \cite{beu}) to reproduce the bump. All the other optical wavelengths agree very well with a rigid shift of the final fitting function for the $R$ band. The results of the fit are reported in Table \ref{tabres}. The peaks in the X-ray and optical bands peak at nearly the same time ($\sim 0.4$~d) suggesting an achromatic bump.

Despite having a different behaviour before the peak there is certainly an additional emission component enhancing the observed flux in the X-ray and in the optical bands. The early time power-law decay ($\alpha_{\rm PL}$) and the rising slopes ($\alpha_{\rm BPL, rise}$) are clearly inconsistent with a single emitting region for the two bands. The early-time X-ray light curve exhibits a steep decay ($\alpha \sim 3.7$) which may arise from the tail-end of the prompt emission, whereas the optical light curve exhibits a much shallower decay. On the contrary, the peak time (t$_{\rm peak}$) of the bump in the light curve at $t \sim 4 \times 10^4$\,s is consistent within errors for both the optical and X-ray bands (however, we note that the decay in the X-ray is slightly steeper than what is observed in the optical). However, optical observations performed at later times with respect to our last optical detection showed that the optical light curve might have undergone a further break to a steeper (consistent with the X-ray decay) value of $\alpha \sim 2.5 \pm 0.2$, suggesting a possible jet collimation of few degrees \cite{urata2}.

 \begin{table*}
 \centering
 \caption[]{Light curve fit results. We model the X-ray and optical light curve with a two component function: an initial power-law (PL) plus a late time broken power-law (BPL). Here we report the initial power-law decay ($\alpha_{\rm PL}$), the rising ($\alpha_{\rm BPL, rise}$), the break time (t$_{\rm break}$), the peak time (t$_{\rm peak}$) and the decaying index ($\alpha_{\rm BPL, decay}$) of the second component. The last column shows the goodness-of-fit (reduced $\chi^2$) and the corresponding degrees of freedom. }
 \label{tabres}
 \begin{tabular}{ccccccc}
 \hline
Band  &  $\alpha_{\rm PL}$ & $\alpha_{\rm BPL, rise}$  &  t$_{\rm break}$ &   t$_{\rm peak}$ & $\alpha_{\rm BPL, decay}$ & $\chi^{2}_{\rm red}$ (d.o.f.)\\
\hline  
 & & & [s] & [s] & &\\
\hline
X-ray & 3.72 $\pm$ 0.08 & -0.38 $\pm$ 0.05 & (6.2 $\pm$ 0.4) $\times$ 10$^{4}$ & (3.2 $\pm$ 0.3) $\times$ 10$^{4}$ & 2.52 $\pm$ 0.11 & 1.20 (238)\\
\hline
Optical & 0.50 $\pm$ 0.05 & -1.53 $\pm$ 0.18 & (3.7 $\pm$ 0.4) $\times$ 10$^{4}$ & (3.6 $\pm$ 0.5) $\times$ 10$^{4}$ & 1.77 $\pm$ 0.11 & 0.95 (37) \\
\hline
\end{tabular}
\end{table*}

\subsection{Spectral energy distribution}

We build the rest-frame spectral energy distribution of the optical afterglow at the post-break time t$^{\rm SED}_{\rm rf}$ = 1 d, corresponding to t$_{\rm obs} = 2.798$ d. The existence of X-ray data at this time allows us to constrain the spectral index ($\beta_{\rm opt-X}$) and the circum-burst absorption ($A_{\rm V}$), under the assumption that the X-ray and optical emission arise from the same spectral component. For these purposes we extracted the X-ray spectrum in the observed time interval [0.4, 2.5] $\times 10^{5}$~s (shaded region left panel Fig. \ref{FigLC}) and re-scaled it to the time selected for the analysis (t$_{\rm SED}$) in order to match the behaviour of the X-ray light curve (see right panel in Fig. \ref{FigLC}).

The data are best described by a single absorbed power-law SMC-model with $\beta_{\rm opt-X} = 0.88 \pm 0.03$ and  $A_{\rm V}^{\rm GRB} = (1.1 \pm 0.3)$ mag (90$\%$ c.l.; $\chi^{2}$/d.o.f. = 31.6/43). Fixing the Galactic column density to N$_{\rm H}^{\rm GAL} = 5 \times 10^{20}$ cm$^{-2}$ we find an X-ray absorbing column density of N$_{\rm H}^{\rm GRB} = (6.6 \pm 0.3) \times 10^{21}$ cm$^{-2}$.

\section{The nature of the late time achromatic peak}

The light curve of GRB\,120326A displays a smooth prolonged re-brightening at late times, both in the optical and X-ray bands (Fig. \ref{FigLC}, left panel). The observed variability might be the result of different processes and can be associated with various forms of late-time energy injection (flare, delayed afterglow onset, refreshed shock emission), with density inhomogeneities in the circum-burst medium or due to some geometrical effect. We now discuss all the possible interpretations of the observed broad peak for GRB\,120326A. 

\subsection{Reverse-shock emission}

A bright (optical) peak could be produced by the reverse shock that propagates back into the shocked material. If present, this peak should happen and be visible in the observed light curve at very early times (t $\le10^{3}$ s) and  subsequently the light curve should display a steep temporal decay index (t$^{-\alpha_{RS}}$ with $\alpha_{RS} \sim 2.0$; \cite{koba1, zhang1}). In the case of GRB\,120326A, the rising light curve after $\sim 5 \times 10^{3}$ s, with $\alpha_{\rm BPL,rise} \sim -1.5$, is preceded by a relatively shallow phase ($\alpha_{\rm PL} \sim 0.5$). At this time the optical light curve has a decay that is too shallow to be associated with a reverse shock and it is probably related to a different component. Moreover no peak at very early times is detected and the only peak visible in the optical light curve happens at t$_{\rm peak} = (3.6 \pm 0.5) \times 10^{4}$~s $\sim$ 0.4 d post-burst, too late to be associated with the reverse-shock emission. We note that the post-bump optical decay index ($\alpha_{\rm BPL, decay} \sim 1.8$) is marginally consistent with the expectation for a reverse shock. All things considered, this explanation is inconsistent with the observed optical light curve of GRB\,120326A.

\subsection{The passage of $\nu_{\rm m}$}

The observed peak could be the signature of the passage of the synchrotron maximal frequency ($\nu_{\rm m}$) across the optical band. If this is the case then the reverse shock is expected to decay as $t^{-0.45}$ (\cite{koba}), while the forward shock will rise as $t^{1/2}$ and decay as $t^{-1}$. In principle this could better explain the initial part of the optical light curve and the temporal difference between the minimum flux reached after $\sim 10^{3}$~s and the optical peak at $\sim 3 \times 10^{4}$~s. This scenario cannot however generate and explain achromatic and coincident peaks at different frequencies, as observed for GRB\,120326A in the optical and in the X-ray bands (Fig. \ref{FigLC}).

\subsection{Onset of the afterglow}

The onset of the forward-shock emission has been observed for many bursts, sometimes at very early times (e.g. GRB~060418, \cite{molinari}) and few times also at later times (e.g. GRB~080129, \cite{greiner}). The multi-wavelength analysis of the very bright GRB\,061007 \cite{mundell, rykoff} showed that the onset of the forward shock can sometimes be inferred to take place at t $\lesssim$ 100~s after the burst event at the optical frequencies. This implies that the reverse shock must have peaked at a typical frequency that is factor $\sim \Gamma_{0}^{2}$ lower with respect to the typical frequency of the optical band. For typical values of $\Gamma_{0}$, if the forward shock peaks in the optical band than the reverse shock will peak at the radio frequencies (low-frequency model, see \cite{koba1} for more details).

As discussed for GRB\,061007 \cite{mundell} and subsequently for GRB\,090313 \cite{mela1} for all the GRBs that display the onset of the optical afterglow it is possible to apply the low-frequency model \cite{koba1} to the observed optical data, making predictions for the expected light curves at the radio frequencies. Since for the case under study in this paper there are few positive detections in the radio-submm band \cite{urata, perley, laskar, staley} we can therefore apply the model to the radio data and understand if what observed in the optical band is the real onset of the forward shock. 

\begin{figure}[|h]
   \centering
   \vspace{-0.3cm}
   \hspace{-0.8cm}
  \includegraphics[width=9.5cm,height=7.0cm]{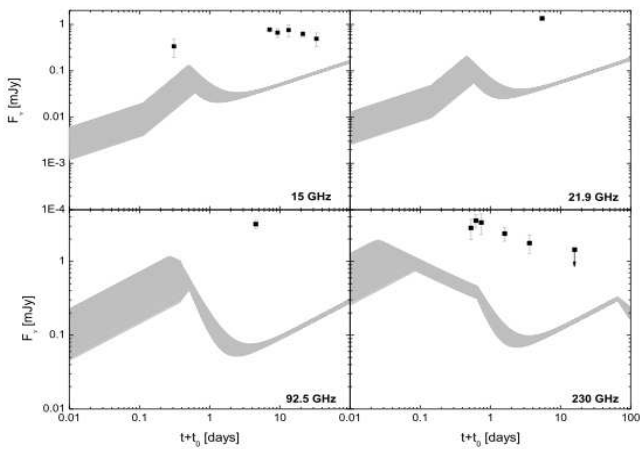}
   \caption{We predicted the light curve in the radio band in the context of the low-frequency model, assuming the late time peak as the possible onset of the forward shock emission. The expected radio light curve is displayed as a grey region that reflects also the uncertainties on the parameters assumed by the model.}
              \label{figradio}
    \end{figure}

In Fig. \ref{figradio}, we compare the confirmed radio detections with the predictions of the low-frequency model (grey region) in the correspondent radio frequencies. The parameters assumed for the model are: $0.1 \leq \epsilon_{\rm e} \leq 0.5$ and $2 \leq p \leq 3$. All the other parameters of the model ($\Gamma_{\rm 0}$, E$_{\rm iso}$, $\epsilon_{\rm B}$, ...) have been calculated from the available data. The predictions, assuming the late peak as the onset of the forward shock, show that the afterglow is expected to peak in the radio band between 0.01 and 1 days after the burst event depending on the frequency, reaching a maximum flux of a few mJy. As seen in Fig. \ref{figradio}, although this scenario does not fully describe the radio observations at high frequencies, the radio predictions for $\nu_{\rm radio} \sim 15$~GHz are only a factor of $\sim 3$ brighter than the model predictions.

A possibility is that the real onset of the afterglow could in fact happen much earlier, without being clearly visible in the optical bands. In fact the observed shallow decay for t $\le 10^{3}$~s could be the decline of a forward shock that peaked at t = t$^{\prime} \le 10^{2}$~s. In that case the average $\Gamma_{\rm peak} \sim187$ and the prediction for the radio emission are even more far off the reported radio detection since the grey regions in Fig. \ref{figradio} will be almost rigidly shifted to the left along the x-axis, as the radio light curve is expected to peak at much earlier times\footnote{The average value for $\Gamma_{\rm peak}$ is calculated using Eq.1 from Molinari et al. (2007) with $n = 1$ cm$^{-3}$. If we assume a radiative efficiency $\eta = 1$ we derive the lower value for the late (early) time observed peak of $\Gamma_{\rm peak} = 18$ ($\geq$ 169), while assuming $\eta = 0.2$ we obtain $\Gamma_{\rm peak} = 22$ ($\geq$ 206). These estimates have been done in the assumption of a homogeneous circum-burst medium; if we assume a wind-like medium ($\rho \sim R^{-2}$) the Lorentz factors for the early and late optical peaks are $\Gamma_{\rm peak}^{wind} \sim 41$ and $\Gamma_{\rm peak}^{wind} \sim 10$, respectively.}. Again, this scenario is incompatible with the radio afterglow detections reported in literature for GRB\,120326A.

\subsection{Late-time optical/X-ray flare emission}

Emission from flares has been detected for many GRBs, superimposed onto their canonical decays from very early times up to $\gtrsim 10^{5}$ s after the burst event \cite{mg}. The observed afterglow variability can be displayed and compared with the kinematically allowed regions in the plane ($\Delta f / f_{\rm peak}$) vs. ($\Delta t / t_{\rm peak}$), describing the increase of the flux with respect the underlying continuum versus the temporal variability \cite{ioka}. 

\begin{figure}
   \centering
   \hspace{-1.0cm}
   \includegraphics[width=7.5cm,height=6.0cm]{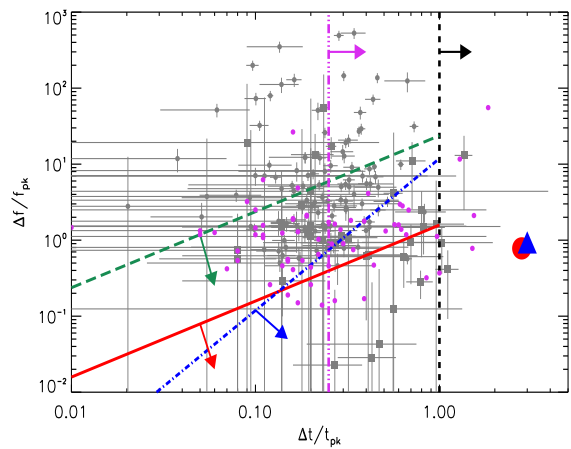}
   \caption{Kinematically allowed regions for afterglow variability in the $\Delta f / f_{\rm peak}$ vs. $\Delta t / t_{\rm peak}$ plane. Solid lines with arrows represent the allowed regions for density fluctuations on-axis (blue), density fluctuations off-axis (red), multiple density fluctuations off-axis (green), refresh shocks (pink) and patchy shell (black), respectively (see \cite{ioka, curran, mg} for details). In this plot we show early (t$_{\rm peak} \le 10^{3}$ s, grey circles; \cite{chinca}) and late (t$_{\rm peak} > 10^{3}$ s, grey squares; \cite{mg}) time X-ray flares, together with UV/optical flares detected at $1\sigma$ confidence level (magenta circles; \cite{swe}). The X-ray and optical peaks observed for GRB\,120326A are shown with a blue triangle and a red circle, respectively.}
              \label{Ioka}
    \end{figure}

As seen in Fig. \ref{Ioka}, all the X-ray flares \cite{chinca, mg}, with the exception of GRB\,050724, can be explained in the context of the internal shocks model, where $\Delta t / t_{\rm peak} < 1$, or single/multiple density fluctuations, if the flux ratio $\Delta f / f_{\rm peak}$ is small. The same conclusions can be drawn in the UV/optical, with the exception of few more events (see Swenson et al. 2013 for details). However, for GRB\,120326A, the observed broad variability at late time in the X-ray and optical bands has $\Delta t / t_{\rm peak} \gg 1$ and therefore cannot be the result of any flare activity. The achromatic bump of GRB\,120326A light curve is more compatible with a possible refreshed shock episode. 

\subsection{Refreshed shock}

Another possibility to explain a late re-brightening in the light curve is to consider a forward shock that is refreshed by a late time energy supply \cite{rees, kumar, sari}. At late times, shells emitted with lower/modest Lorentz factor ($\Gamma \sim 10 \div 20$) catch up with faster shells ($\Gamma \geq 100$ ) that have been already slowed down by the interaction with the external material, injecting energy into the afterglow shock and causing a significant re-brightening in the observed light curve. This scenario has been successfully invoked to explain, for example, the numerous bumps of the optical light curve of GRB\,030329 \cite{granot}. 

Under the simplified assumption of only two shells colliding, with $\Gamma_{\rm fast} \geq 10^{2}$ and $\Gamma_{\rm slow} = \Gamma_{\rm 0} \sim 10$, it can be shown (see \cite{genet}) that the two shells will collide at a time

\vspace{-0.5cm}
\begin{eqnarray}
t_{\rm shock} \approx 1.66~E_{\rm \gamma,iso,53}^{1/3}~n^{-1/3}~\Gamma_{\rm slow, 10}^{-8/3} ~~~~{\rm days}
\end{eqnarray}

\noindent where E$_{\rm \gamma,iso,53}$ is the isotropic energy in units of 10$^{53}$ erg, $n$ is the density of the external medium (assumed to be = 1 cm$^{-3}$ for uniform medium) and $\Gamma_{\rm slow, 10}$ is the Lorentz factor of the slow shell in units of ten. In the case under study E$_{53} \sim$ 0.35, and in order to explain the bump happening at $t = t_{\rm shock} \sim 0.4$~days the slow material adding energy into the forward shock should have a $\Gamma_{\rm slow, 10} = \Gamma_{\rm 0} / 10 \sim 2$. In fact, as shown in Sec. 4.3, the average values of $\Gamma_{\rm peak}$ estimated for the late broad peak of GRB\,120326A is of the order of a few tens and therefore the refreshed-shock scenario could explain the observed behaviour.

\begin{figure}
 \centering
  \includegraphics[width=5.5cm,height=7.5cm,angle=270]{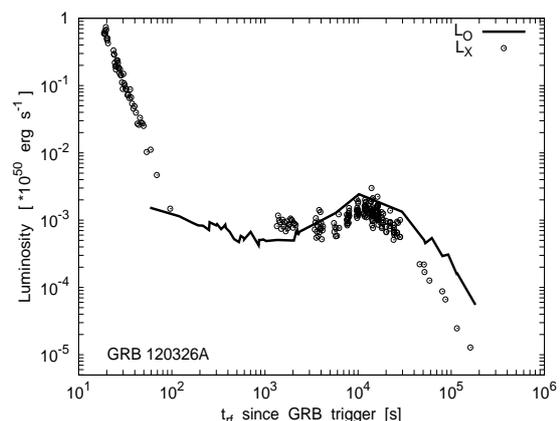}
 \caption{Rest-frame X-ray (open circles) and optical (solid line) luminosity. At late times the achromatic behaviour of the light curve is evidence supporting the interpretation of a possible refreshed shock or geometrical effect.}
          \label{FigLumOX1}%
\end{figure}

A possible drawback for this interpretation is that 1) the peak that we see at late time is probably not the onset of the forward shock and therefore the estimate of $\Gamma_{\rm peak}$ could not be done accurately and 2) the observed rising and decaying indices ($\alpha_{\rm BPL, rise}$ and $\alpha_{\rm BPL, decay}$ reported in Table 2) in the optical and X-ray bands seem to slightly differ while these values are expected to be consistent in the two bands. However, this can be explained by the different contributions of the early time emission observed in the two bands. As can be see in Fig. \ref{FigLC}, the X-ray emission for $t < 10^{3}$~s is very steep and its contribution to the second component would be more relevant for the rising part of the X-ray emission after $\sim 3 \times 10^{3}$~s, while it would be negligible at very late times. Instead, in the optical band the early emission is very flat and it will still contribute at later times, making the light curve in that band flatter. A marginal difference seems to be present when looking to the observed light curves at particular frequencies, probably due to the different time coverage (Fig. \ref{FigLC}). When considering the rest-frame X-ray and optical luminosities (Fig. \ref{FigLumOX1}), however, the agreement between these two bands is straightforward. Despite these small differences the refreshed-shock scenario cannot be excluded at high confidence level.

The energy injection scenario has also been analysed in details for the wind-like and homogeneous circumburst medium by Hou et al. (2014). The authors found that a stellar wind circumburst environment could provide a reasonable fit of the observed X-ray and optical light curves of GRB\,120326A \cite{hou}.

\subsection{Density Bump}

An increase of the external medium density is sometimes invoked to explain the late time re-brightening in the optical light curve when a corresponding bump is not seen in the X-ray band \cite{lazzati, dai}. A sharp and large jump in a uniform density profile is however needed in order to produce an observable increase in the observed light curves. A sudden enhancement by factor $a$=10 ($\geq 10^{2}$) of the medium density will correspond to a variation of $\Delta \alpha \leq 0.4$ ($\approx 1.0$) in the observed temporal slopes. Those variations in the temporal decay are relevant only for the radio frequencies. In the optical band, even a large density enhancement would generate small hard-to-detect variations \cite{granot, eerten}. In general, if the observed bump in the optical light curve for GRB\,120326A is produced by the external shock then any density variation is unlikely to produce the achromatic signature observed in Fig. \ref{FigLC}. Therefore, this interpretation seems unlikely.

\subsection{Geometrical effect}

For a jet with an opening angle $\theta_{\rm jet}$ of a few degrees, an observer located at an angle $\theta_{\rm view} > \theta_{\rm jet}$ will see a bump in the light curve at relatively late times when, due to the jet deceleration, the bulk Lorentz factor of the jet is $1/\Gamma \sim \theta_{\rm view}-\theta_{\rm jet}$ (i.e. \cite{granot2, guidorzi3}). First order calculations suggest that a jet with a typical opening angle of $\sim$ 3 degrees seen at 5 degrees off--axis could account for the peak. In this scenario, which requires further modelling of the optical and X-ray afterglow light curve, and whose detailed study is left for a future work, the peak should be achromatic as observed in GRB\,120326A. However, a jet seen off axis cannot account for the relatively high energy of the prompt emission observed in this burst. Even a jet with typical parameters as those described above and an isotropic equivalent energy as large as 10$^{54}$~erg could not account for the observed isotropic energy observed since the de-beaming factor would be $\propto \delta^2$. 

A possible solution is that a wider jet component or structure (e.g. jet tails) are present and these intercept the line-of-sight so as to account for both the prompt emission energy (and peak energy) and the emission observed in the optical before the bump.



\subsection{Gravitational microlensing}

In principle, achromatic fluctuations in GRB afterglow light curves, observed less than one day post-burst, could be the result of gravitational microlensing \cite{loeb}. Such an effect would produce a magnification of the observed flux adding, for the X-ray and optical ($R$) band, a sharp peak on the declining part of the afterglow light curve. This explanation was successfully applied to GRB\,000301C, a GRB at redshift $z = 2.04$ that displayed an achromatic bump 3.8 days after the burst. That bump corresponded to a flux (magnitude) magnification of $\sim 2$ ($\sim 1$~mag) in its light curve \cite{garna}. 

In the case under study we can estimate the lower limit of flux magnification factor at the peak time $\mu_{obs}$(t$_{\rm peak}) \sim 23.5$, as the ratio between the maximum of the light curve shown in Fig. \ref{FigLC} and the flux of the early power-law component extrapolated to t = t$_{\rm peak}$. The estimated factor correspond to a magnitude magnification of at least 3.4 mag in the afterglow light curve. Such a strong magnification factor, coupled with the broadness of the observed peak observed, makes this interpretation very unlikely for GRB\,120326A.

\section{Conclusions}

Our multi-band analysis of GRB\,120326A allows us to conclude that the striking feature observed in the late-time afterglow light curve, the broad achromatic re-brightening, cannot be ascribed to reverse- or forward-shock emission, nor to the passage of the synchrotron frequency through the optical band. The long duration and magnitude of the re-brightening also make the late flare, the gravitational microlensing and the density bump origin inconsistent with the observed data. 

Although the light curves are not all fully sampled across all wavebands, we have established that: 1) in the optical/IR bands the available data are consistent with an achromatic behaviour, with all light curves described by the same fitting function rigidly shifted at different wavelengths; 2) the pre-bump (different) emission observed in the X-ray and optical bands is not simply explained with a single emission component and an additional contribution must be present that shapes differently the observed light curves, after the end of the prompt emission; and 3) the observed late-time behaviour could be explained either by a late-time refreshed forward shock (prolonged energy supply from the central engine) or by a geometrical effect (a two-component jet seen slightly off-axis). We cannot favour one of these two scenarios over the other. Further detailed study of the multi-wavelengths light curves for a larger sample of events is needed, coupled with good theoretical predictions to compare with well sampled observations.

\begin{acknowledgements}
The research activity of AM and MGB is supported by ASI grant INAF I/004/11/1. The Liverpool Telescope is operated by Liverpool John Moores University at the Observatorio del Roque de los Muchachos of the Instituto de Astrof\'{i}sica de Canarias.  CGM acknowledges support from the Royal Society, the Wolfson Foundation and the Science and Technology Facilities Council. DK work is partially supported by Optical $\&$ Near-Infrared Astronomy Inter-University Cooperation Program, the MEXT of Japan. This work made use of data supplied by the UK Swift Science Data Centre at the University of Leicester.
\end{acknowledgements}


\end{document}